\documentclass[conference]{ieeeconf}

\IEEEoverridecommandlockouts                              

\usepackage{xr}

\newcommand{\omitted}[1]{}

\title{\fontsize{16.4}{16.4} \selectfont 
Leveraging Untrustworthy Commands for Multi-Robot Coordination in Unpredictable Environments: A Bandit Submodular Maximization Approach
}
\author{Zirui Xu,$^\star$ Xiaofeng Lin,$^\star$ Vasileios Tzoumas
	\thanks{$^\star$Contributed equally to this work.}
	\thanks{
 Z. Xu and V. Tzoumas are with the Department of Aerospace Engineering, University of Michigan, Ann Arbor, MI 48109 USA;  {\tt\footnotesize \{ziruixu, vtzoumas\}@umich.edu}} 
 	\thanks{X. Lin is with the Division of Systems Engineering, Boston University, Boston, MA 02215 USA; {\tt\footnotesize xfl@bu.edu}} 
}

\usepackage{cite}

\usepackage{comment}
\usepackage{siunitx}
\usepackage{relsize}
\usepackage{ifthen}
\usepackage[colorinlistoftodos]{todonotes}

\usepackage[vlined,ruled,linesnumbered]{algorithm2e}
\usepackage{graphics} 
\usepackage{rotating}
\usepackage{color}
\usepackage{enumerate}
\usepackage[T1]{fontenc}
\usepackage{psfrag}
\usepackage{epsfig} 
\usepackage{booktabs}
\usepackage{graphicx,url}
\usepackage{multirow}
\usepackage{array}
\usepackage{latexsym}
\usepackage{amsfonts}
\usepackage{amsmath}
\usepackage{amssymb}
\usepackage{xstring}
\usepackage{algorithmic}
\usepackage{multirow}
\usepackage{xcolor}
\usepackage{prettyref}
\usepackage{flexisym}
\usepackage{bigdelim}
\usepackage{breqn} 
\usepackage{listings}
\let\labelindent\relax
\usepackage{xspace}
\usepackage{bm}
\graphicspath{{./figures/}}
\usepackage{tikz}
\usetikzlibrary{matrix,calc}
\usepackage{lipsum}
\usepackage{mdwlist}

\let\itemize\undefined
\makecompactlist{itemize}{stditemize}
\usepackage{enumitem}
\usepackage{caption}
\usepackage{amsthm}
\usepackage{mathtools}

\makeatletter
\let\NAT@parse\undefined
\makeatother
\usepackage{hyperref}
\usepackage{cleveref}

\hypersetup{%
  colorlinks=true,
  linkcolor=blue,
  filecolor=magenta,      
  urlcolor=black,
  citecolor=red,
  linkbordercolor={0 0 1}
}

\newtheorem{theorem}{Theorem}
\newtheorem{problem}{Problem}

\newtheorem{definition}{Definition}
\newtheorem{proposition}{Proposition}

\newcommand{\bdmath}{\begin{dmath}}
\newcommand{\edmath}{\end{dmath}}
\newcommand{\beq}{\begin{equation}}
\newcommand{\eeq}{\end{equation}}
\newcommand{\bdm}{\begin{displaymath}}
\newcommand{\edm}{\end{displaymath}}
\newcommand{\bea}{\begin{eqnarray}}
\newcommand{\eea}{\end{eqnarray}}
\newcommand{\beal}{\beq \begin{array}{lll}}
\newcommand{\eeal}{\end{array} \eeq}
\newcommand{\beas}{\begin{eqnarray*}}
\newcommand{\eeas}{\end{eqnarray*}}
\newcommand{\ba}{\begin{array}}
\newcommand{\ea}{\end{array}}
\newcommand{\bit}{\begin{itemize}}
\newcommand{\eit}{\end{itemize}}
\newcommand{\ben}{\begin{enumerate}}
\newcommand{\een}{\end{enumerate}}

\newcommand{\calA}{{\cal A}}
\newcommand{\calB}{{\cal B}}

\newcommand{\calN}{{\cal N}}

\newcommand{\calT}{{\cal T}}

\newcommand{\calV}{{\cal V}}

\definecolor{myblue}{RGB}{65 105 225}

\newcommand{\hide}[1]{}

\newcommand{\hiddenText}{{\color{gray} hidden text.}}
\newcommand{\hideWithText}[1]{\hiddenText}

\newcommand{\opt}{^{\star}}

\newcommand{\scenario}[1]{{\fontsize{9}{9}\selectfont\sf #1}\xspace}

\newcommand{\ie}{\emph{i.e.},\xspace}
\newcommand{\eg}{\emph{e.g.},\xspace}
\newcommand{\myin}{\, \in \,}

\newcommand{\sg}{\scenario{SG}}
\newcommand{\banalg}{\scenario{BSG}}

\newcommand{\myParagraph}[1]{{\bf #1.}\xspace}

\renewcommand{\opt}{\scenario{OPT}}

\newcommand{\expsix}[1]{\scenario{EXP3$^{\star}$-SIX$|_{#1}$}}

\newcommand{\solbsg}{\calA^{\banalg}}
\newcommand{\solmeta}{\calA^{\sgmeta}}

\newcommand{\solopt}{\calA^{\opt}}

\newcommand{\sgmeta}{\scenario{MetaBSG}}

\newcommand{\expthree}{\scenario{EXP3$^{\star}$-SIX}}
\newcommand{\expthreeix}{\scenario{EXP3-IX}}
\newcommand{\human}{\scenario{ExtComm}}

\newcommand{\distfsf}{p}

\newcommand{\distmeta}{p}

\newcommand{\acthuman}{a^{\human}}

\PassOptionsToPackage{end}{algorithmic}

\begin{document}

\maketitle

\thispagestyle{empty}
\pagestyle{empty}

\begin{abstract}%
We study the problem of multi-agent coordination in unpredictable and partially-observable environments with \textit{untrustworthy external commands}. The commands are actions suggested to the robots, and are untrustworthy in that their performance guarantees, if any, are unknown.  Such commands may be generated by human operators or machine learning algorithms and, although untrustworthy, can often increase the robots' performance in complex multi-robot tasks. We are motivated by complex multi-robot tasks such as target tracking, environmental mapping, and area monitoring. Such tasks are often modeled as submodular maximization problems due to the information overlap among the robots. We provide an algorithm, \textit{Meta Bandit Sequential Greedy} (\sgmeta), which enjoys performance guarantees even when the external commands are arbitrarily bad. \sgmeta leverages a meta-algorithm to learn whether the robots should follow the commands or a recently developed submodular coordination algorithm, \textit{Bandit Sequential Greedy} (\banalg)~\cite{xu2023bandit}, which has performance guarantees even in unpredictable and partially-observable environments. Particularly, \sgmeta asymptotically can achieve the better performance out of the commands and the \banalg algorithm, quantifying its suboptimality against the optimal time-varying multi-robot actions in hindsight. Thus, \sgmeta can be interpreted as \textit{robustifying the untrustworthy commands}. We validate our algorithm in simulated scenarios of multi-target tracking.
\end{abstract}

\section{Introduction}\label{sec:Intro}

The future of autonomy will involve multiple robots collaborating on complex tasks, such as {target tracking}~\cite{tokekar2014multi}, {environmental modeling}~\cite{krause2008near}, and {area coverage}~\cite{xu2022resource}. Such tasks can be modeled as maximization problems of the form
\begin{equation}\label{eq:intro}
	\max_{a_{i,\,t}\,\in\,\mathcal{V}_i,\,  \forall\, i\,\in\, \calN}\
	f_t(\,\{a_{i,\,t}\}_{i\myin \calN}\,), \ \ t=1,2,\ldots,
\end{equation}
where $\calN$ is the robot set, $a_{i,\,t}$ is robot $i$'s action at time step $t$, $\calV_i$ is robot $i$'s set of available actions, and $f_t:2^{\prod_{i \in \calN}\calV_i}\mapsto\mathbb{R}$ is the objective function that captures the task utility at time step $t$. The optimization problem in~\cref{eq:intro} is generally NP-hard~\cite{Feige:1998:TLN:285055.285059} but near-optimal approximation algorithms are possible in polynomial time when $f_t$ is \textit{submodular}~\cite{fisher1978analysis} ---submodularity is a diminishing returns property capturing the potential information overlap among the robots. The seminal approximation algorithm for~\cref{eq:intro} with $f_t$ being submodular is the \textit{Sequential Greedy} (\sg) algorithm~\cite{fisher1978analysis} that achieves a $1/2$-approximation bound. A wide range of multi-robot information-gathering tasks including all above can be modeled as submodular coordination, and thus \sg-based algorithms are commonly applicable~\cite{krause2008near,singh2009efficient,tokekar2014multi,atanasov2015decentralized,gharesifard2017distributed,grimsman2018impact,corah2021scalable,schlotfeldt2021resilient,liu2021distributed,robey2021optimal,konda2022execution,xu2022resource}.

\begin{figure}[t]
    \captionsetup{font=footnotesize}
    \centering
    \includegraphics[width=.98\columnwidth]{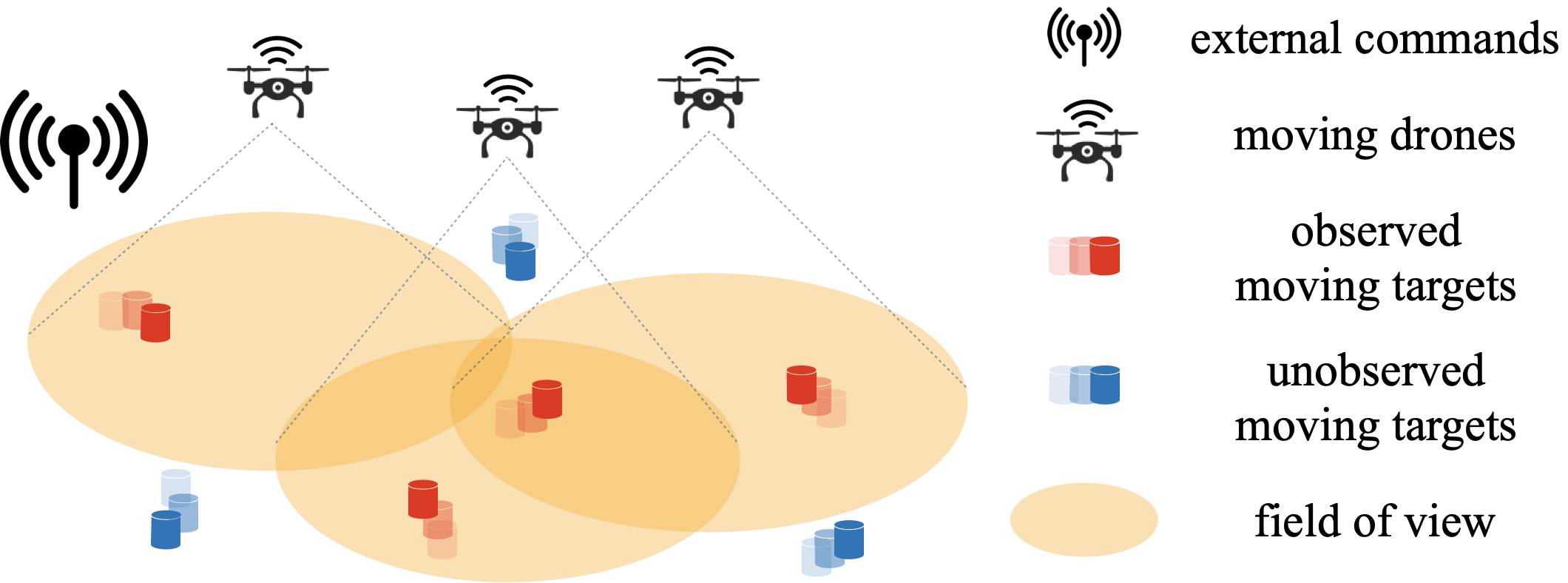}\\
    \caption{\textbf{Multi-Robot Coordination in Unpredictable and Partially Observable Environments with Untrustworthy External Commands: Target Tracking.} In unpredictable environments, the robots would benefit from machine learning algorithms or human operators that can suggest actions (commands) based on predictions about the future.  But such suggested actions are untrustworthy in that they also often rely on erroneous predictions, lacking performance guarantees.  For example, in target tracking, human operators may sometimes be able to learn the targets’ behavior and suggest commands that can maximize the number of tracked targets; but at other times the suggested commands can be highly ineffective when the targets are able to change the pattern of their behavior. In this paper, we propose an online coordination framework to robustly leverage untrustworthy commands, guaranteeing bounded suboptimality even when the external commands are arbitrarily bad.
   }\label{fig:partial}
   \vspace{-12mm}
\end{figure}

But the applicability of the said algorithms becomes limited in \textit{unpredictable} and \textit{partially-observable} environments where at each time step $t$ the objective function $f_t$ is unknown a priori and only partially known a posteriori. For example, in target tracking (Fig.~\ref{fig:partial}), drones are often tasked to coordinate their motion to maximize at each time step the number of tracked targets.  But in unpredictable and partially-observable environments, (i) the robots may be unaware of the targets' intentions and motion model, thus being unable to plan effective actions by simulating the future, \ie being unable to know $f_t$, and (ii) the robots may carry sensors with a limited field of view, thus being unable to reason even in hindsight whether alternative actions could have been more effective in tracking targets, \ie a higher $f_t$ value.  Such limited access to information is called \textit{bandit feedback}~\cite{lattimore2020bandit}.

Although algorithms have been developed for multi-robot coordination in unpredictable environments with bandit feedback, these algorithms either (i) offer performance guarantees but can be overly conservative, or (ii) offer no performance guarantees but can be highly effective. For example, the recently developed algorithm \textit{Bandit Sequential Greedy} (\banalg)~\cite{xu2023bandit} quantifies its suboptimality with respect to the optimal time-varying multi-robot actions in hindsight, but it can be overly conservative since it assumes that the environment can evolve arbitrarily, even adversarially. Similarly, machine learning algorithms can be highly effective in predicting the evolution of the environment but offer no formal performance guarantees; specifically, they may perform near-optimally when the future environment resembles the training samples, but they perform arbitrarily bad otherwise~\cite{sunderhauf2018limits}.\footnote{Other works that consider unpredictable and partially-observable environments are~\cite{baykal2017persistent,zhang2019partially,lee2021upper,landgren2021distributed,dahiya2022scalable,wakayama2023active} but their focus is on linear objectives, instead of submodular, not accounting for the information overlap among the agents.}$^,$\footnote{Additional algorithms~\cite{streeter2008online,streeter2009online,suehiro2012online,golovin2014online,clark2014distributed,chen2018online,zhang2019online,chen2020black} have been proposed with guaranteed bounded suboptimality for tasks where $f_t$ is unknown a priori, but these algorithms assume $f_t$ is \textit{static} instead of time-varying, and require continuous optimization techniques that are more expensive in computation and communication resources in the multi-agent coordination settings~\cite{xu2022resource}.}

\myParagraph{Goal} In this paper, we aim to prove an algorithm for bandit submodular coordination that guarantee bounded suboptimality in unpredictable and partially-observable environments even when the external commands are arbitrarily bad.

\myParagraph{Contributions} 
We provide an algorithm for the problem of multi-agent coordination in unpredictable and partially-observable environments with {untrustworthy external commands} (\Cref{subsec:meta}).  We name the algorithm \textit{Meta-Bandit Sequential Greedy} (\sgmeta). \sgmeta is based on the classic framework of multi-arm bandits along with the recently developed \banalg algorithm for bandit submodular coordination in unpredictable and partially-observable enviroments~\cite{xu2023bandit}. \sgmeta has the properties 
(\Cref{sec:guarantees}):
\begin{itemize}[leftmargin=9pt]
	\item \textit{Computational Complexity}: For each agent $i$, \sgmeta requires only $1$ function evaluation and $O(\log{T})$ additions and multiplications per agent per round.
	\item \textit{Approximation Performance}: \sgmeta enjoys performance guarantees even when the external commands are arbitrarily bad.  Specifically, \sgmeta  asymptotically matches the better out of the external commands and the \banalg algorithm. As such, \sgmeta leverages ``good'' commands to improve \banalg's performance, and robustifies ``bad'' commands per \banalg's suboptimality against the optimal time-varying multi-robot actions in hindsight.  Particularly, if the commands have an approximation ratio close to $1$, then \sgmeta will also achieve a performance close to the optimum. To this end, \sgmeta leverages a multi-armed bandit algorithm to learn whether the robots should follow the external commands or the \banalg algorithm.
\end{itemize}

\myParagraph{Numerical Evaluations} 
We evaluate \sgmeta in simulated scenarios of target tracking with multiple robots, where the robots carry noisy sensors with limited field of view to observe the targets (\Cref{sec:experiments}).  We consider scenarios where (i) $2$ robots pursue $2$ targets while receiving suboptimal external commands, and (ii) $2$ robots pursue $4$ targets while receiving near-optimal external commands. The simulations validate that \sgmeta enables the robots to learn the better strategy out of the commands and the \banalg algorithm. 

\section{Bandit Submodular Coordination with Untrustworthy External commands}\label{sec:problem}

We define the problem \textit{Bandit Submodular Coordination with Untrustworthy External Commands}. 

To this end, we use the notation:
\begin{itemize}
    \item $[T]\triangleq\{1,\dots,T\}$ for any positive integer $T$;
    \item $f(\,a\,|\,\calA\,)\triangleq f(\,\calA \cup \{a\}\,)-f(\,\calA\,)$ is the marginal gain of set function $f:2^\calV\mapsto \mathbb{R}$ for adding $a \in \calV$ to $\calA \subseteq\calV$.
    \item $|\calA|$ is the cardinality of a discrete set $\calA$. 
\end{itemize}

The following preliminary framework is considered, which is introduced in \cite{xu2023bandit} with the exception of the untrustworthy external commands.

\myParagraph{Agents}  $\calN$ is the set of all agents, and the terms ``agent'' and ``robot'' are used interchangeably. The agents coordinate their actions to complete a task, and they can observe each other's selected actions at each time step to achieve this goal.

\myParagraph{Actions} $\calV_i$ is a \textit{discrete} and \textit{finite} set of actions always available and known a priori to robot $i$.  For instance, $\calV_i$ may comprise motion primitives~\cite{tokekar2014multi} or discrete control inputs~\cite{atanasov2015decentralized} that robot $i$ can execute to move in the environment. 

\begin{figure*}[t!]
    \captionsetup{font=footnotesize}
    \centering
    \includegraphics[width=0.9\linewidth]{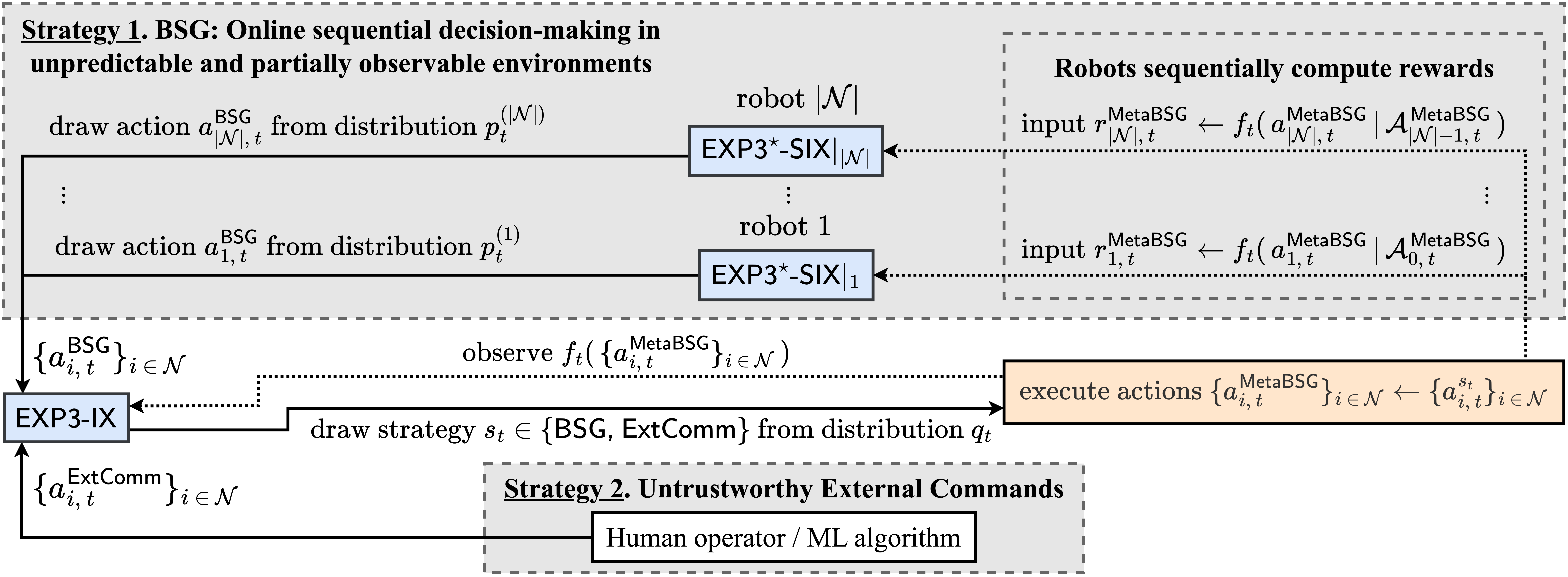}\\
    \caption{\textbf{Block Diagram of {\smaller \sf MetaBSG}.} {\smaller \sf MetaBSG} is a meta-algorithm that decides at each time step whether to follow the online coordination algorithm {\smaller \sf BSG} (Strategy 1) which has a conservative performance guarantee, or some untrustworthy external commands (Strategy 2) which may outperform BSG. 
    At each time step $t$, first, the subroutine {\smaller \sf BSG} generates candidate actions $\{a_{i,\,t}^{\textsf{BSG}}\}_{i\,\in\,\mathcal{N}}$ by drawing from the distributions $p_{t}^{(1)},\dots,p_{t}^{(|\calN|)}$. Then, {\smaller \sf MetaBSG} receives the untrustworthy external commands $\{a_{i,\,t}^{\textsf{ExtComm}}\}_{i\,\in\,\mathcal{N}}$ and uses the subroutine {\smaller \sf EXP3-IX} to decide whether to follow $\{a_{i,\,t}^{\textsf{BSG}}\}_{i\,\in\,\mathcal{N}}$ or $\{a_{i,\,t}^{\textsf{ExtComm}}\}_{i\,\in\,\mathcal{N}}$ by drawing from distribution $q_t$. Afterwards, the robots execute actions $\{a_{i,\,t}^{\textsf{MetaBSG}}\}_{i\,\in\,\mathcal{N}}$ and observe the $f_t$ values of their actions. Finally, {\smaller \sf EXP3-IX} updates the distribution $q_{t}$ and {\smaller \sf BSG} updates the distributions $p_{t}^{(1)},\dots,p_{t}^{(|\calN|)}$ for the next time step $t+1$.
    \vspace{-5mm}
    }\label{fig:block-diagram}
\end{figure*}

\myParagraph{Objective Function with Submodular Structure} 
The agents coordinate their actions $\{a_{i,\,t}\}_{i\myin \calN}$ to maximize $f_t(\{a_{i,\,t}\}_{i\myin \calN})$, the objective function at time $t$. $f_t(\cdot)$ captures the multi-robot task's utility and is time-varying across $t\in[T]$. In Fig.~\ref{fig:partial}, for example, the objective function equals $4$ since four targets are within the field of view of the robots given the robots' and targets' positions at time $t$.

In information-gathering multi-robot tasks
such as target tracking~\cite{tokekar2014multi}, environmental modeling~\cite{krause2008near}, and area coverage~\cite{xu2022resource}, common objective functions quantify the area/information covered due to the robots' actions. They satisfy the properties defined below (Definition~\ref{def:submodular}).
\begin{definition}[Normalized and Non-Decreasing Submodular Set Function{~\cite{fisher1978analysis}}]\label{def:submodular}
A set function $f:2^\calV\mapsto \mathbb{R}$ is \emph{normalized and non-decreasing submodular} if and only if 
\begin{itemize}
\item (Normalization) $f(\,\emptyset\,)=0$;
\item (Monotonicity) $f(\,\calA\,)\leq f(\,\calB\,)$, $\forall\,\calA\subseteq \calB\subseteq \calV$;
\item (Submodularity) $f(\,s\,|\,\calA\,)\geq f(\,s\,|\,{\mathcal{B}}\,)$, $\forall\,\calA\subseteq {\mathcal{B}}\subseteq\calV$ and $s\in \calV$.
\end{itemize}
\end{definition}

Normalization holds without loss of generality, while monotonicity and submodularity are inherent properties of the function. For the example in Fig.~\ref{fig:partial}, let $f(\mathcal{A})$ represent the number of targets tracked by a set $\mathcal{A}$ of drones. Intuitively, the monotonically non-decreasing property states that deploying more drones will not result in tracking fewer targets. Additionally, the submodularity property implies that the marginal increase in tracked targets due to deploying a drone $s$ decreases as more drones are already deployed.

\myParagraph{Environment} The agents coordinate in an unpredictable and partially-observable environment where at each time step $t\in[T]$, $f_t(\cdot)$ is unknown to the agents, and, even after time $t$, $f_t(\cdot)$ becomes only partially known, as explained in more detail in the next example and the paragraph below. For example, in the multi-target tracking scenario depicted in Fig.~\ref{fig:partial}, where the robots have \underline{no} model of the future motion of targets, it is impossible at time $t-1$ for the robots to predict the targets' location at time $t$; moreover, even after the robots have executed their actions at $t$, their collaborative field of view covers only the red targets instead of all targets including the blue ones. 

\myParagraph{Bandit Feedback} The agents can only receive \textit{bandit feedback}~\cite{lattimore2020bandit} because of the unpredictable and partially observable environment: (i) $f_t(\cdot)$ is unknown to the agents prior to time $t$, and (ii) even after the execution of actions $\{a_{i,\,t}\}_{i\myin \calN}$, the agents can only evaluate $f_t(\,\calA\,)$ for all $\calA\subseteq\{a_{i,\,t}\}_{i\myin \calN}$ due to partial observability. Had the agents been able to observe the whole environment after executing their actions, they could evaluate in hindsight $f_t(\,\calA\,)$ for all $\calA\subseteq\{\calV_i\}_{i\myin \calN}$, \ie the performance of any actions that they could have chosen for time $t$, and this type of feedback is \textit{full-information feedback}~\cite{cesa2006prediction}. 

\myParagraph{Untrustworthy External Commands} At each step $t$, the robots have the option to execute the actions $\{a_{i,\,t}^\human\}_{i\myin\calN,\,t\myin[T]}$ given by external commands. For example, the commands may be generated by {human operators} or {machine learning algorithms}~\cite{sunderhauf2018limits}. The commands are untrustworthy in that their near-optimality is unknown.  Particularly, the suboptimality of $\{a_{i,\,t}^\human\}_{i\myin\calN,\,t\myin[T]}$ with respect to the optimal actions $\{a_{i,\,t}^\opt\}_{i\myin\calN,\,t\myin[T]}$ is
\begin{equation}\label{eq:external_bound}
    \sum_{t=1}^{T} f_t(\,\{\acthuman_{i,\,t}\}_{i\myin\calN}\,)\, \geq \,\beta\,\sum_{t=1}^{T} f_t(\,\{a_{i,\,t}^\opt\}_{i\myin\calN}\,),
\end{equation}where $\beta$ is unknown. When $\beta=1$, then the commands are optimal; otherwise, they are suboptimal ---\eg $\beta=0$ means the commands offer no performance guarantee.

\myParagraph{Problem Definition}  In this paper, we focus on:
\begin{problem}[Bandit Submodular Coordination with Untrustworthy External Commands]
\label{pr:online}
Assume a time horizon of operation discretized to $T$ time steps. At each time step $t\in[T]$, the agents $\calN$ receive external commands $\{\acthuman_{i,\,t}\}_{i\myin\calN}$, and then select actions $\{a_{i,\,t}\}_{i\myin \calN}$ \emph{online} to solve
\begin{equation}
    \max_{a_{i,\,t}\,\in\,\mathcal{V}_i, \,
         \forall\, i\,\in\, \calN}\
    f_t(\,\{a_{i,\,t}\}_{i\myin \calN}\,),
\end{equation}
where $f_t: 2^{\prod_{i\myin \calN} \,\calV_i}\mapsto \mathbb{R}$ is a normalized and non-decreasing submodular set function, and the agents $\calN$ become aware of the values of $f_t(\,\calA\,)$ \emph{only once} they have selected $\{a_{i,\,t}\}_{i\myin \calN}$, $\forall\,\calA\subseteq\{a_{i,\,t}\}_{i\myin \calN}$.
\end{problem}

At each time step $t$, \Cref{pr:online} requires the robots to collaboratively  choose actions $\{a_{i,\,t}\}_{i\myin \calN}$ to maximize $f_t$, based on  the {history of past commands} and \textit{rewards} ---{reward at a time step $t$ is the value  $f_t(\,\{a_{i,\,t}\}_{i\myin \calN}\,)$}.  Particularly, \Cref{pr:online} requires finding an algorithm $\phi$ such that 
$$\{a_{i,\,t}\}_{i\myin\calN} = \phi\left(
\begin{tabular}{@{}l@{}}
	{\small all agents' received rewards till $t-1$;} \\
    {\small all agents' external commands till $t-1$;}\\
    {\small all agents' received external commands at $t$}
\end{tabular}\right),$$where $\phi$ decides at each $t$ whether the robots should adopt as actions the external commands.  Ideally, $\phi$ would guarantee bounded suboptimality even when the external commands are arbitrarily bad, \ie even when $\beta=0$.  That is, $\phi$ would guarantee that there exists a $b\in(0,1]$ such that
\begin{equation}\label{eq:desired_bound}
    \sum_{t=1}^{T} f_t(\,\{a_{i,\,t}\}_{i\myin\calN}\,)\, \geq \,\max\{b, \beta\}\,\sum_{t=1}^{T} f_t(\,\{a_{i,\,t}^\opt\}_{i\myin\calN}\,),
\end{equation}
despite being oblivious to $\beta$ and despite the bandit feedback.
In this paper, we propose \sgmeta for the role of $\phi$.


\section{Meta Bandit Sequential Greedy (\sgmeta) Algorithm} \label{sec:algorithm}

We present the algorithm for \Cref{pr:online}, \textit{Meta Bandit Sequential Greedy} (\sgmeta) (Fig.~\ref{fig:block-diagram}). \sgmeta learns whether the robots should choose actions per \banalg or per the external command.  The problem thus has the form of a \textit{two-armed bandit problem}, where the one arm is {to follow the untrustworthy external commands \human}, and the other arm is {to follow \banalg}. \sgmeta employs the \expthreeix algorithm~\cite{neu2015explore} to solve this two-armed bandit problem. We next first present how \banalg solves the vanilla \textit{bandit submodular coordination} problem (\Cref{subsec:tracking-best-sequence}) where there are no external commands, then present how \expthreeix solves the two-armed bandit problem (\Cref{subsec:exp3ix}), and finally present the \sgmeta (\Cref{subsec:meta}). 

\subsection{\banalg Algorithm}\label{subsec:tracking-best-sequence}
We first present the \banalg algorithm \cite{xu2023bandit} (\Cref{alg:online}) for solving the vanilla bandit submodular coordination problem where there are no external commands (\Cref{pr:bandit-submodular-coordination}). 

We use the notation:
\begin{itemize}
    \item the optimal sequence of actions in hindsight, \ie the optimal sequence of actions the agents would select if they fully knew $f_1,\dots,f_T$ a priori, is denoted by {\vspace{-3mm}\small$$\{a_{i,\,t}^\opt\}_{i\in \calN,\,t\in [T]}\in\arg\max_{a_{i,\,t}\in\mathcal{V}_{i},\, i\in\calN}\,\sum_{t=1}^T f_t(\{a_{i,\,t}\}_{i\in\calN});\vspace{-2mm}$$}
    \item ${\bf 1}(\cdot)$ is the indicator function, \ie ${\bf 1}(x)=1$ if the event $x$ is true, otherwise ${\bf 1}(x)=0$.
    \item $\Delta(T) \triangleq \sum_{t=1}^{T-1} \sum_{i\myin \calN} \;{\bf 1}(\,a_{i,\,t}^\opt\neq a_{i,\;t+1}^\opt\,)$ is the environment's adversarial effect that evaluates how many times the agents' optimal actions shift over $[T]$.
\end{itemize}

\begin{problem}[Bandit Submodular Coordination\cite{xu2023bandit}]
\label{pr:bandit-submodular-coordination}
At each time $t\in[T]$, the agents select actions $\{a_{i,\,t}\}_{i\myin \calN}$ \emph{online} to
\begin{equation}
    \max_{a_{i,\,t}\,\in\,\mathcal{V}_i, \,
         \forall\, i\,\in\, \calN}\
    f_t(\,\{a_{i,\,t}\}_{i\myin \calN}\,),
\end{equation}
where $f_t: 2^{\prod_{i\myin \calN} \,\calV_i}\mapsto \mathbb{R}$ is a normalized and non-decreasing submodular set function,  and the agents become aware of the values of $f_t(\,\calA\,)$ \emph{only once} they have selected $\{a_{i,\,t}\}_{i\myin \calN}$, $\forall\,\calA\subseteq\{a_{i,\,t}\}_{i\myin \calN}$.
\end{problem}

\banalg approximates a solution to \Cref{pr:bandit-submodular-coordination} by employing (i) a sequential coordination scheme to enable the robots coordinate the selection of their actions ---the sequential coordination scheme is inspired by the Sequential Greedy algorithm~\cite{fisher1978analysis}--- and employing (ii)  an online learning algorithm on each robot that enables the robot to ``guess'' which a near-optimal action at each time $t$, despite not knowing $f_t$; the online learning algorithm onboard each robot is \expthree, presented in~\cite{xu2023bandit}. In more detail, \banalg starts by instructing each agent $i\in \calN$ to initialize \expsix{i}, an instance of \scenario{EXP3$^\star$-SIX} (line 2). Then, at each time step $t\in[T]$, given the probability distribution $\distfsf_t^{(i)}$ output by \expsix{i}, the agents each draw an action $a_{i,\,t}^\banalg$ and execute them (lines 3-6), {without getting any information about $f_t$ yet}. Afterwards, each agent $i$ sequentially receives from agent $i-1$ $\solbsg_{i-1,\,t}$, the actions of all agents $\{1,\dots,i-1\}$, and then observes $f_t(\,\solbsg_{i,\,t}\,)$ (lines 7-10).
Finally, each agent $i$ computes $r_{i,\, t}^\banalg$, the reward (marginal gain) of $a_{i,\,t}^\banalg$ given $\solbsg_{i-1,\,t}$, and inputs $r_{i,\, t}^\banalg$ to \expsix{i} per line 8 of~\cite[Algorithm 1]{xu2023bandit} (lines 11-14). \expsix{i} will use this input to compute $\distfsf_{t+1}^{(i)}$, the probability distribution over the agent $i$'s actions for the next time step. {While the sequential ordering of the agents may matter the actual performance, the theoretical approximation performance guarantee of \banalg is not affected, which is shown as follows. }

\setlength{\textfloatsep}{3mm}
\begin{algorithm}[t]
	\caption{\mbox{Bandit Sequential Greedy (\banalg)~\cite{xu2023bandit}}
	}
	\begin{algorithmic}[1]
		\REQUIRE Time steps $T$ and agents' action sets $\{\mathcal{V}_i\}_{i \myin \calN}$.
		\ENSURE \!Actions $\{a_{i,\, t}^\banalg\}_{i\,\in\,\calN}$ at each $t\in[T]$.
		\medskip
		\STATE Order the agents in $\calN$ such that $\calN=\{1,\dots,|\calN|\}$;
		\STATE Each agent $i\in\calN$ initializes an instance of \scenario{EXP3$^\star$-SIX}, \expsix{i}, with time step $T$ and action set $\calV_i$;
		\FOR {\text{each time step} $t\in [T]$}
        \STATE \textbf{receive} $\{\distfsf_t^{(i)}\}_{i\myin\calN}$                    from \mbox{$\{$\expsix{i}$\!\!\}_{i\myin\calN}$}; 
		\STATE \textbf{draw} actions $\{a_{i,\,t}^\banalg\}_{i\,\in\,\calN}$ from $\{\distfsf_t^{(i)}\}_{i\,\in\,\calN}$;
			\STATE \textbf{execute} $\{a_{i,\, t}^\banalg\}_{i\,\in\,\calN}$; 
			\STATE $\solbsg_{0,\,t} \gets \emptyset$;
			\FOR {$i = 1, \dots, |\calN|$} 
			\STATE $\solbsg_{i,\,t} \gets \solbsg_{i-1,\,t}\cup \{a_{i,\,t}^\banalg\}$;
			\STATE \textbf{observe} $f_t(\,\solbsg_{i,\,t}\,)$;
			\STATE $r_{i,\, t}^\banalg\gets f_t(\,a_{i,\,t}^\banalg \,|\, \solbsg_{i-1,\,t}\,)$;
			\STATE \textbf{input} $r_{i,\, t}^\banalg$ to \expsix{i} (line 8 of \cite[Algorithm 1]{xu2023bandit});
			\ENDFOR
		\ENDFOR
	\end{algorithmic}\label{alg:online}
\end{algorithm}

\begin{theorem}[Performance Guarantee of \banalg~\cite{xu2023bandit}]\label{lem:bsg} For \Cref{pr:bandit-submodular-coordination}, \banalg guarantees
{\small
\begin{align}\label{aux:tracking_sec}
    &\mathbb{E}\left[\sum_{t=1}^{T}f_t(\,\{a_{i,\,t}^\banalg\}_{i\myin\calN}\,)\right] \geq \\\nonumber
    &\frac{1}{2}\,\mathbb{E}\left[\sum_{t=1}^{T}f_t(\,\{a_{i,\,t}^\opt\}_{i\myin\calN}\,)\right] - \tilde{O}\left[\sqrt{T\Delta(T)}\,\right] - \log{\left(\frac{1}{\delta}\right)}\,\tilde{O}\left[\sqrt{T}\right]
\end{align}}with probability at least $1-\delta$, where $\delta\in(0,1)$ is the confidence level, $\tilde{O}(\cdot)$ hides $\log$ terms, and the expectation is due to \banalg's internal randomness.\footnote{{We have the $\mathbb{E}\left[\sum_{t=1}^{T}f_t(\{a_{i,\,t}^\opt\}_{i\in\calN})\right]$ with respect to {\smaller \sf BSG}'s internal randomness  in eq.~\eqref{aux:tracking_sec} because $f_t$ can be adaptive to the actions selected by {\smaller \sf BSG} up to time $t-1$ and, thus, the optimal actions in hindsight $\{a_{i,\,t}^\opt\}_{i\myin\calN}$ are also a function of the actions selected by {\smaller \sf BSG} up to $t-1$.}}
\end{theorem}Therefore, if $\Delta(T)$ is sublinear in $T$, then $\Delta(T)/T\rightarrow 0$ for $T\rightarrow \infty$, and \banalg will enable the agents to asymptotically adapt to coordinate as if they knew $f_1,\dots,f_{T}$ a priori, matching the $1/2$-approximate performance of Sequential Greedy. {$\Delta(T)$ being sublinear in $T$ holds in environments where, \eg the evolution is unknown yet predefined, instead of being adaptive to the agents’ actions~\cite{xu2023bandit,xu2023online}.  Then, $\Delta(T)$ is uniformly bounded since increasing the discretization density of time horizon $H$, \ie increasing the number of time steps $T$, does not affect the environment's evolution. This result agrees with the intuition that the agents should be able to adapt to an unknown but non-adversarial environment when they re-select actions with high enough frequency.} 

\subsection{\expthreeix Algorithm}\label{subsec:exp3ix}
We present \expthreeix (\Cref{alg:expthreeix}), the meta-algorithm that \sgmeta employs to solve a two-armed bandit problem (\Cref{pr:strategy-selection}), such that \sgmeta's performance over $[T]$ matches up to ``good'' commands and is no worse than \banalg.

\begin{problem}[Bandit Problem with Two Arms: External Commands and \banalg]\label{pr:strategy-selection}
For a time horizon of operation discretized to $T$ time steps, at each time step $t\in[T]$, the agents are given $\{a^{\human}_{i,\,t}\}_{i\myin\calN}$ and $\{a^{\banalg}_{i,\,t}\}_{i\myin\calN}$, and then select $s_t\in\{\human, \banalg\}$  \emph{online} to solve
\begin{equation}
    \max_{s_{t}\myin\{\human,\,\banalg\},\, t\myin[T]} \;\; \sum_{t=1}^T\;f_t(\,\{a^{s_t}_{i,\,t}\}_{i\myin\calN}\,),
\end{equation}where the agents become aware of $f_t(\,\{a^{s_t}_{i,\,t}\}_{i\myin\calN}\,)\in[0,1]$ \emph{only once} they have chosen $s_{t}$.
\end{problem} 

We present how \expthreeix~\cite{neu2015explore} (\Cref{alg:expthreeix}) solves \Cref{pr:strategy-selection}. 
\Cref{alg:expthreeix} maintains two weight parameters $z_{\human,t}$ and $z_{\banalg,t}$ for the untrustworthy commands and \banalg (lines 1-2). At each time step $t\in[T]$,  \Cref{alg:expthreeix} first computes the output probability distribution $q_t$ based on $z_{t}$, then receives the new reward $r_{s_t,\,t}$ of selecting strategy $s_t$ based on $q_t$ (lines 3-5). Using $r_{s_t,\,t}$, \Cref{alg:expthreeix} updates the estimate $\tilde{r}_t$ for the two strategies' rewards (lines 6-7). $\tilde{r}_t$ is an optimistically biased estimate of $r_{s_t,\,t}$, \ie larger than the unbiased estimate, so as to encourage the algorithm to explore the less selected strategy~\cite{neu2015explore,kocak2014efficient}. Finally, \Cref{alg:expthreeix} updates $z_{t+1}$ for time $t+1$ (lines 8-9). The higher $\eta$ is, the faster \Cref{alg:expthreeix} adapts to the better strategy. Per \cite[Theorem~1]{neu2015explore}, \Cref{alg:expthreeix} guarantees,
\vspace{-1mm}
    {\small\begin{align}\label{eq:bound_expthreeix}
        &\hspace{-.7cm}\mathbb{E}\left[\sum_{t=1}^{T}f_t(\,\{a^{s_t}_{i,\,t}\}_{i\myin\calN}\,)\right] \\
        &\geq\max_{s\myin\{\human,\,\banalg\}}{\mathbb{E}\left[\sum_{t=1}^{T} f_t(\,\{a^{s}_{i,\,t}\}_{i\myin\calN}\,)\right]}  \nonumber\\\nonumber
        &\qquad - 4\sqrt{T\log{2}} - \left(\sqrt{4T/\log{2}}+1\right) \log{(2/\delta)},\vspace{-2mm}    
    \end{align}}with probability at least $1-\delta$, for any $\delta \in (0,1)$, where the expectation is due to \Cref{alg:expthreeix}'s internal randomness.
\setlength{\textfloatsep}{3mm}
\begin{algorithm}[t]
	\caption{\expthreeix~\cite{neu2015explore} for Bandit Problem with Two Arms: External Commands and \banalg
	}\label{alg:expthreeix}
	\begin{algorithmic}[1]
		\REQUIRE Time steps $T$ and strategy set $\{\human, \banalg\}$.
		\ENSURE Probability distribution $q_t \myin \{[0,1]^{2}:\|q_t \|_1=$ $1\}$ over $\{\human, \banalg\}$ at each $t\in[T]$.
		\smallskip
		
		\STATE $\eta\gets\sqrt{{\log{2}}\,/\,{T}}$; $\gamma=\eta/2$;
		\STATE $z_{1}\gets\left[z_{\human,\,1}, z_{\banalg,\,1}\right]^\top$ with $z_{\human,\,1}=z_{\banalg,\,1}=1$;
        \FOR {each time step $t\in[T]$}
        \STATE \textbf{output} $q_t\gets{z_t}\,/\,{\|z_t\|_1}$;
        \STATE \textbf{receive} reward $r_{s_t,\,t}\in [0,1]$  of selecting strategy  $s_t\in\{\human, \banalg\}$ at $t$; 
        \STATE $\Tilde{r}_{s,\,t} \gets 1 - \frac{{\bf 1}(s_t\,=\,i)}{q_{i,\,t}\,+\,\gamma}(1\,-\,r_{s_t,\,t})$, for both $s\in\{\human, \banalg\}$;
        \STATE $\Tilde{r}_{t}\gets\left[\Tilde{r}_{\human,\,t},\,\Tilde{r}_{\banalg,\,t}\right]^\top$;
        \STATE $z_{\human,\,t+1}\gets z_{\human,\,t}\exp{(\,\eta \,\Tilde{r}_{\human,\,t}\,/\, \|\Tilde{r}_t\|_1)}$; $z_{\banalg,\,t+1}\gets z_{\banalg,\,t}\exp{(\,\eta \,\Tilde{r}_{\banalg,\,t}\,/\, \|\Tilde{r}_t\|_1)}$;
		\ENDFOR
	\end{algorithmic}
\end{algorithm}

\subsection{\sgmeta Algorithm}\label{subsec:meta}

\setlength{\textfloatsep}{3mm}
 \begin{algorithm}[t]
 	\caption{\mbox{Meta Bandit Sequential Greedy (\sgmeta)}
 	}
 	\begin{algorithmic}[1]
		\REQUIRE Time steps $T$ and agents' action sets $\{\mathcal{V}_i\}_{i \myin \calN}$. 
		\REQUIRE Actions $\{\acthuman_{i,\,t}\}_{i\myin\calN}$ from \human and $\{a_{i,\, t}^\banalg\}_{i\,\in\,\calN}$ from \banalg at each $t\in[T]$.
 		\ENSURE Actions $\{a_{i,\, t}^\sgmeta\}_{i\,\in\,\calN}$ at each $t\in[T]$.
 		\medskip
  		\STATE Order the agents in $\calN$ such that $\calN=\{1,\dots,|\calN|\}$;
		\STATE Each agent $i\in\calN$ initializes an instance of \scenario{EXP3$^\star$-SIX}, \expsix{i}, with time step $T$ and action set $\calV_i$;

 		\FOR {\text{each time step} $t\in [T]$}
            \STATE \textbf{receive} $\{\distfsf_t^{(i)}\}_{i\myin\calN}$ from \mbox{$\{$\expsix{i}$\!\!\}_{i\myin\calN}$}; 
			\STATE \textbf{draw} actions $\{a_{i,\,t}^\banalg\}_{i\,\in\,\calN}$ from $\{\distfsf_t^{(i)}\}_{i\,\in\,\calN}$;
            \STATE \textbf{receive} actions $\{\acthuman_{i,\,t}\}_{i\myin\calN}$;
            \STATE \textbf{receive} $q_t$ from \expthreeix (line 4 of \Cref{alg:expthreeix});
   			\STATE \textbf{draw} a strategy $s_t\in\{\human, \banalg\}$ from $q_t$; 
 			\STATE \textbf{execute} $\{a_{i,\, t}^\sgmeta\}_{i\,\in\,\calN}\gets\{a_{i,\, t}^{s_t}\}_{i\,\in\,\calN}$; 
 			\STATE $\solmeta_{0,\,t} \gets \emptyset$;
 			\FOR {$i = 1, \dots, |\calN|$} 
 			\STATE $\solmeta_{i,\,t} \gets \solmeta_{i-1,\,t}\cup \{a_{i,\,t}^\sgmeta\}$;
 			\STATE \textbf{observe} $f_t(\,\solmeta_{i,\,t}\,)$;
 			\STATE $r_{i,\, t}^\sgmeta\gets f_t(\,a_{i,\,t}^\sgmeta \,|\, \solmeta_{i-1,\,t}\,)$;
 			\STATE \textbf{input} $r_{i,\, t}^\sgmeta$ to \expsix{i} (line 8 of \cite[Algorithm 1]{xu2023bandit})
			\ENDFOR
   \STATE \textbf{input} $f_t(\,\{a_{i,\, t}^\sgmeta\}_{i\,\in\,\calN}\,)$ to \expthreeix (line 5 of \Cref{alg:expthreeix});
 		\ENDFOR
 	\end{algorithmic}\label{alg:sgmeta}
 \end{algorithm}

We present \sgmeta (\Cref{alg:sgmeta}) to enable the agents to decide online whether to follow the untrustworthy external commands or \banalg; the challenge is that the agents are unaware of whether the commands are near-optimal. To this end, \sgmeta embeds \expthreeix into \banalg to decide which strategy the agents should follow at each $t\in[T]$, among the external commands or \banalg, such that the agents' performance over $[T]$ \mbox{can match the better strategy of the two.}

\sgmeta's process differs from \banalg's in that, instead of directly executing the actions $\{a_{i,\,t}^\banalg\}_{i\,\in\,\calN}$ given by \banalg, \sgmeta decides which strategy to follow based on a probability distribution $q_t$ given by \expthreeix. In more detail, \sgmeta starts by instructing each agent $i$ to initialize \expsix{i} (line 2).  
Then, at each time step $t\in[T]$: 
\begin{itemize}
    \item[i)] Each agent $i$ draws an action $a_{i,\,t}^\banalg$ given the probability distribution $\distmeta_t^{(i)}$ output by \expsix{i} and receives external command $a_{i,\,t}^\human$ (lines 3-6).
    \item[ii)] The agents choose a strategy based on the probability distribution $q_t$ given by \expthreeix (lines 7-8).
    \item[iii)] The agents execute actions $\{a_{i,\, t}^\sgmeta\}_{i\,\in\,\calN}$ (line 9).
    \item[iv)] Each agent $i$ receives from agent $i-1$ $\solmeta_{i-1,\,t}$, the actions of all previous agents,  and then observes $f_t(\,\solmeta_{i,\,t}\,)$ (lines 10-13).
    \item[v)] Each agent $i$ computes the reward $r_{i,\, t}^\sgmeta$, \ie the marginal gain of $a_{i,\,t}^\sgmeta$ given $\solmeta_{i-1,\,t}$, and inputs $r_{i,\, t}^{\sgmeta}$ to \expsix{i} per line 8 of \cite[Algorithm 1]{xu2023bandit} (lines 13-14). \expsix{i} will use $r_{i,\, t}^\sgmeta$ to update the probability distribution $\distmeta_{t+1}^{(i)}$ over the action set $\calV_i$ for $t+1$ (lines 14-16). 
    \item[vi)] Finally, the algorithm inputs $f_t(\,\{a_{i,\, t}^\sgmeta\}_{i\,\in\,\calN}\,)$ to \expthreeix per line 5 of \Cref{alg:expthreeix}, and \expthreeix will use $f_t(\,\{a_{i,\, t}^\sgmeta\}_{i\,\in\,\calN}\,)$ to update the probability distribution $q_t$ over the two strategies for $t+1$  (lines 17-18).
\end{itemize}
We can see that except for steps (ii) and (vi) that maintain an \expthreeix, \sgmeta has the same process as \banalg.


\section{Performance Guarantees of \sgmeta}\label{sec:guarantees}
We present the computational complexity of \sgmeta (\Cref{subsec:complexity}) and its approximation performance with untrustworthy external command (\Cref{subsec:guarantee}). 

\subsection{Computational Complexity of \sgmeta}\label{subsec:complexity}

\begin{proposition}[Computational Complexity]\label{pror:computations}
\sgmeta requires each agent $i\in\calN$ to perform $T$ function evaluations and $O(T\log{T})$ additions and multiplications over $T$ rounds.
\end{proposition}

At each $t\in[T]$, \sgmeta requires each agent to perform $1$ function evaluation to compute the reward (\Cref{alg:online}'s line 14) and $O(\log{T})$ additions and multiplications to run \expsix{i} \cite[Algorithm 1]{xu2023bandit}. In addition, $O(1)$ additions and multiplications are required for running \expthreeix.

\subsection{Approximation Performance of \sgmeta}\label{subsec:guarantee}

We bound \sgmeta's suboptimality with respect to the optimal sequence of actions the agents' would select if they knew $\{f_t\}_{t\myin[T]}$ a priori (\Cref{th:main}). 
We prove that the performance guarantee of \sgmeta asymptotically matches the better out of \banalg and the external commands. As such, \sgmeta manages to leverage ``good'' commands to improve the approximation bound and robustify ``bad'' commands by using \banalg to lower-bound itself. 
Particularly, if the commands 
have an approximation ratio close to $1$, then \sgmeta will also achieve a performance close to the optimum. 

We use the notation:
\begin{itemize}
    \item $\solopt_t\triangleq\{a_{i,\,t}^\opt\}_{i\,\in\, \calN}$ is the agents $\calN$'s optimal in hindsight actions at $t\in[T]$ as defined in \Cref{subsec:tracking-best-sequence};
    \item $\{a_{i,\, t}^\sgmeta\}_{i\,\in\,\calN}$ is the agents $\calN$'s actions given by \sgmeta at $t\in[T]$.
    \item $\Delta(T) \triangleq \sum_{t=1}^{T-1} \sum_{i\myin \calN} \;{\bf 1}(\,a_{i,\,t}^\opt\neq a_{i,\;t+1}^\opt\,)$ is the environment's adversarial effect that evaluates how many times the agents' optimal actions shift over $[T]$.
\end{itemize}

\begin{theorem}[Approximation Performance of \sgmeta]\label{th:main} \sgmeta instructs the agents in $\calN$ to select online actions $\{\solmeta_t\}_{t\myin[T]}$ that guarantee
{\small\begin{align}\label{eq:main_bound}
    &\mathbb{E}\left[\sum_{t=1}^{T} f_t(\,\solmeta_t\,)\right] \geq \\
    &\!\!\!\max\Bigg\{\overbrace{\frac{1}{2}\,\mathbb{E}\left[\sum_{t=1}^{T}f_t(\,\calA_t^\opt\,)\right] - \tilde{O}\left[\sqrt{T\Delta(T)}\,\right] - \underbrace{\log{\left(\frac{1}{\delta}\right)}\,\tilde{O}\left[\sqrt{T}\right]}_{\textbf{sublinear}}}^{\banalg\textbf{'s approximation bound}}, \nonumber\\\nonumber
    &\!\overbrace{\beta\,\mathbb{E}\left[\sum_{t=1}^{T}f_t(\,\calA_t^\opt\,)\right]}^{\human\textbf{'s approximation bound}}\Bigg\} - \underbrace{O\left[\sqrt{T}\right] - \log{\left(\frac{1}{\delta}\right)}\,O\left[\sqrt{T}\right]}_{\textbf{sublinear}},
\end{align}}holds with probability at least $1-\delta$, for any $\delta \in (0,1)$, and the expectation is due to \sgmeta's internal randomness. 
\end{theorem}

\Cref{th:main} bounds the suboptimality of \sgmeta as a function of the approximation bounds of \banalg and \human. \sgmeta asymptotically matches the better out of \banalg and the external commands.  Specifically, as the number of time steps $T$ increases, since the last term in \cref{eq:main_bound} is sublinear, if $\{\calA_{t}^\human\}_{t\myin[T]}$ has a better approximation guarantee than $\{\solbsg_{t}\}_{t\myin[T]}$ over $[T]$, {\ie if $\beta >1/2$,} then \sgmeta's performance will follow the external command, becoming $\beta$-approximate as $T\rightarrow\infty$. Otherwise, if $\{\calA_{t}^\human\}_{t\myin[T]}$ are worse than $\{\solbsg_{t}\}_{t\myin[T]}$, \sgmeta's performance will follow \banalg. All in all, if $\Delta(T)$ is sublinear in $T$, then $\Delta(T)/T\rightarrow 0$ for $T\rightarrow \infty$, and thus \banalg will asymptotically match the $1/2$-approximate performance of Sequential Greedy~\cite{fisher1978analysis}, and thus \sgmeta will asymptotically become $\max{(1/2, \beta)}$-approximate as $T\rightarrow\infty$.


\vspace{-2mm}
\section{Numerical Evaluation in Multi-Target Tracking Tasks with Multiple Robots}\label{sec:experiments}

\begin{figure}[b!]
    \captionsetup{font=footnotesize}
    \centering
    \includegraphics[width=0.61\columnwidth]{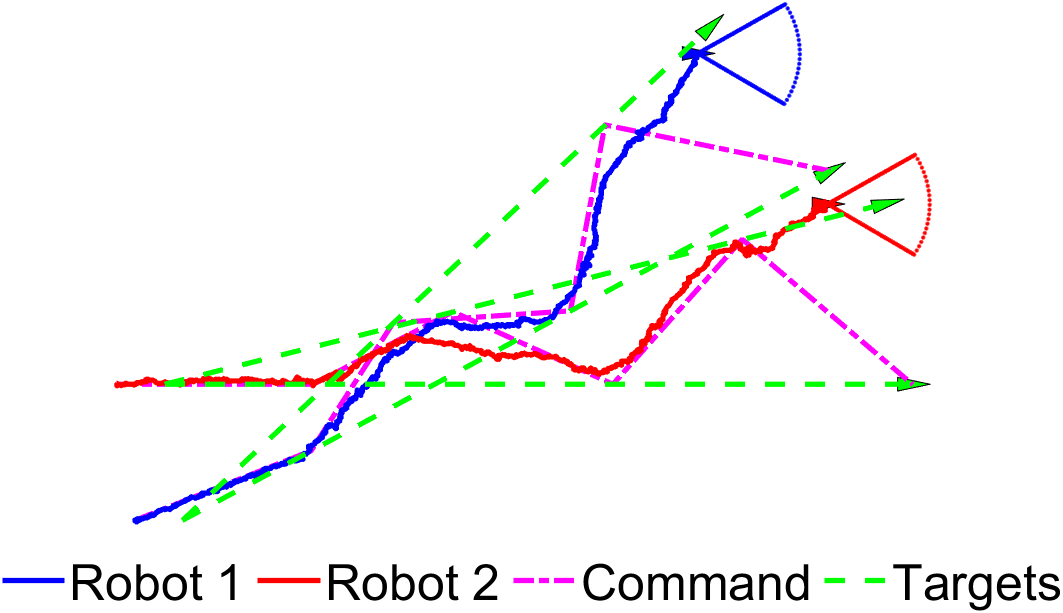}\\
    \caption{\textbf{Target-Tracking Instance of $2$ Robots Tracking $4$ Targets.} The robots and targets all move on the same 2-D plane. A robot can collect a target's range and bearing measurements but only when the target is inside the robot's field of view. The measurements are assumed corrupted by unbiased Gaussian noises whose variance increases for a further target. The targets' motion models are unknown, hence the robots can only coordinate only based on the observations so far.  Yet, the robots have the option at any time step to follow external commands that aim to guide the robots. However, the commands' near-optimality is also unknown.  Thus,
    the robots have to decide online whether to (i) follow the commands or to (ii) ignore them and coordinate nevertheless based on the observations so far. 
    }\label{fig:partial-sim}
\end{figure}

We evaluate \sgmeta in simulated scenarios of target tracking with multiple robots, where the robots carry noisy sensors with limited field of view to observe the targets. The robots have the option to either follow \banalg or follow untrustworthy external commands whose near-optimality is unknown a priori. External commands are generated before each simulation. Particularly, we consider scenarios where (i) $2$ robots pursue $2$ targets while receiving suboptimal external commands, and (ii) $2$ robots pursue $4$ targets while receiving near-optimal external commands.

\myParagraph{Benchmark Algorithm} 
We compare \sgmeta's performance with \banalg and the external commands.

\vspace{1mm}
\myParagraph{Common Simulation Setup across Simulated Scenarios} 

\paragraph{Targets} The targets $\calT$ move on a 2D plane.   We introduce the targets' motion model within each particular scenario considered in \Cref{subsec:evaluation}.

\paragraph{Robots}
The robots move in the same 2D environment as the targets. To move in the environment, each robot $i\in\calN$ can perform one of the actions  $\calV_i\triangleq$ \{``upward'', ``downward'', ``left'', ``right'', ``upleft'', ``upright'', ``downleft'', ``downright''\}  at a constant speed.

\paragraph{Sensing} 
We consider that each robot $i$ can collect range and bearing measurements about the targets' position inside its field of view. After selecting actions $\{a_{i,\,t}\}_{i\myin\calN}$ at time $t$, the robots share their measurements with one another, enabling each robot $i$ to have an estimate of $d_t(a_{i,\,t},\,j)$, the distance from robot $i$ to target $j$, given that $j$ is observed by a robot as a result of actions $\{a_{i,\,t}\}_{i\myin\calN}$. 

\newcommand{\introFigTitleWidth}{0.2cm}
\newcommand{\introFigColWidth}{4.5cm}
\newcommand{\introFigSpacing}{\hspace{-2mm}}
\newcommand{\intoFigNameSpacing}{}
\newcommand{\advFigColWidth}{7cm}

\begin{figure*}[t!]
    \captionsetup{font=footnotesize}
	\begin{center}
	\hspace{-9cm}
      \begin{minipage}{\columnwidth}
            \begin{tabular}{p{\introFigTitleWidth}p{\introFigColWidth}p{\introFigColWidth}p{\advFigColWidth}}%
            \begin{minipage}{\introFigTitleWidth}%
            \end{minipage}
            &   
            \begin{minipage}{\introFigColWidth}%
                  \centering
                  \rotatebox{0}{{\sf \smaller\textbf{MetaBSG}}\vspace{-4cm}}
            \end{minipage}
            &
            \begin{minipage}{\introFigColWidth}%
                  \centering
                  \rotatebox{0}{{\sf \smaller\textbf{BSG}}\vspace{-4cm}}
            \end{minipage}
            &            
            \begin{minipage}{\advFigColWidth}%
                  \centering
                  \rotatebox{0}{\sf \smaller \textbf{MetaBSG vs. BSG vs. Command}}
            \end{minipage}
            \\
            \hspace{-3mm}\begin{minipage}{\introFigTitleWidth}%
                  \rotatebox{90}{\begin{tabular}{c}
                       \textbf{2 vs.~2 } \\
                       \textbf{w/ Suboptimal Commands}
                  \end{tabular} \hspace{-0.5cm}}
            \end{minipage}
            &
            \begin{minipage}{\introFigColWidth}%
                \vspace{1.5mm}
                  \centering%
                  \includegraphics[width=.66\columnwidth]{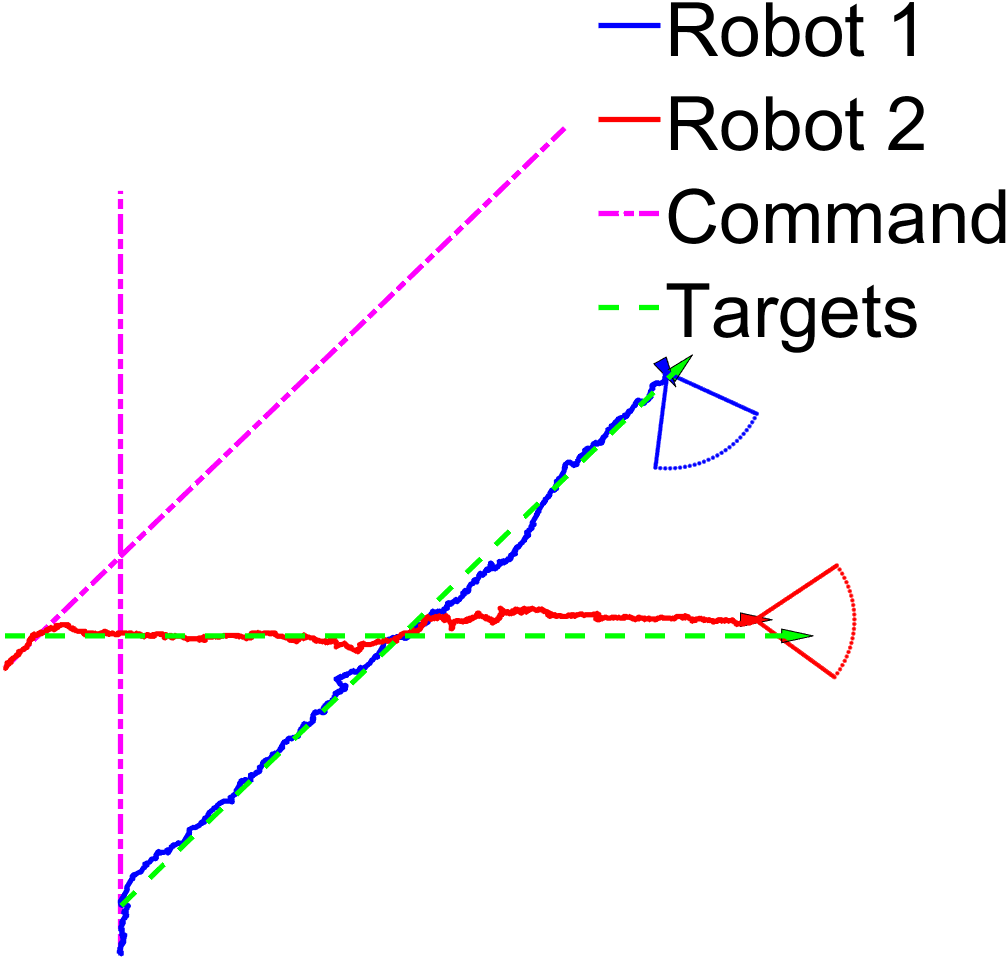} \\
                  \vspace{-0.6mm}
                  \caption*{(a) {\smaller \sf MetaBSG}: 2 robots and 2 targets.
                  }
            \end{minipage}
            &
            \begin{minipage}{\introFigColWidth}
                  \centering%
                  \vspace{7mm}
                  \includegraphics[width=.6\columnwidth]{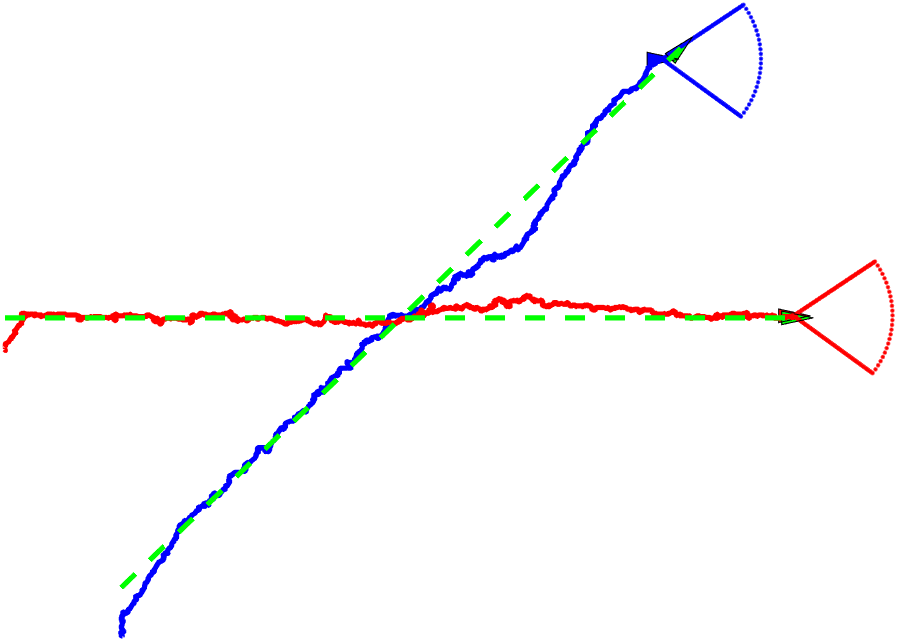} \\ 
                   \vspace{2.9mm}
                  \caption*{(b) {\smaller \sf BSG}: 2 robots and 2 targets.
                  }
            \end{minipage}            
            &
            \begin{minipage}{\advFigColWidth}%
                  \centering%
                  \hspace*{-1mm}
                  \includegraphics[width=\columnwidth]{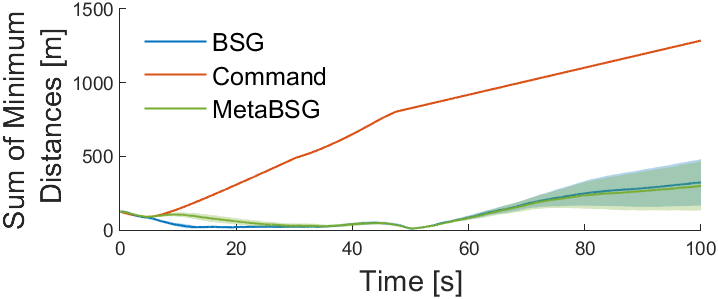} \\
                  \caption*{
                  (c) {\smaller \sf MetaBSG} vs. {\smaller \sf BSG} vs. {\smaller \sf Command}:
                  2 robots and 2 targets.
                  }
            \end{minipage}
            \\
            \hspace{-3mm}\begin{minipage}{0.3cm}%
                  \rotatebox{90}{\begin{tabular}{c}
                       \textbf{2 vs.~4 } \\
                       \textbf{w/ Optimal Commands}
                  \end{tabular} \hspace{0cm}}
            \end{minipage}
            &
            \begin{minipage}{\introFigColWidth}%
            \vspace{5mm}
            \centering%
            \includegraphics[width=.8\columnwidth]{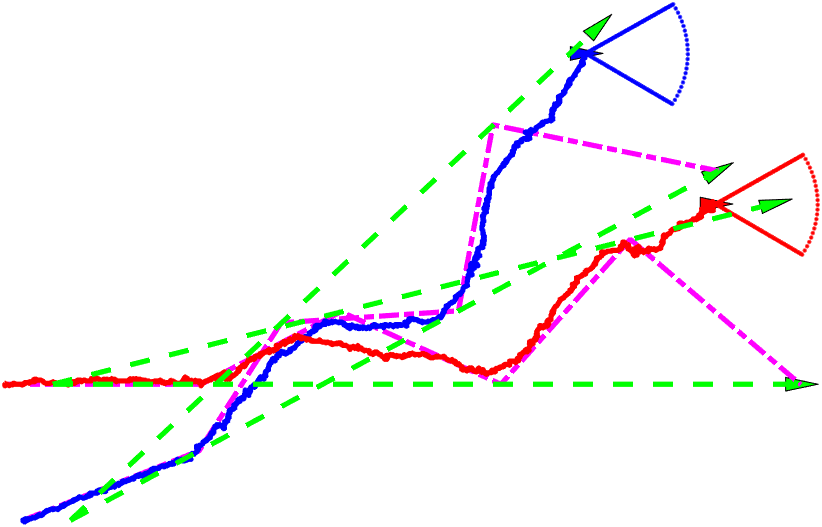} \\
            \vspace{2.6mm}
            \caption*{(d) {\smaller \sf MetaBSG}: 2 robots and 4 targets.
            }
            \end{minipage}
            &
            \begin{minipage}{\introFigColWidth}
                  \centering%
            \vspace{5mm}
            \includegraphics[width=.79\columnwidth]{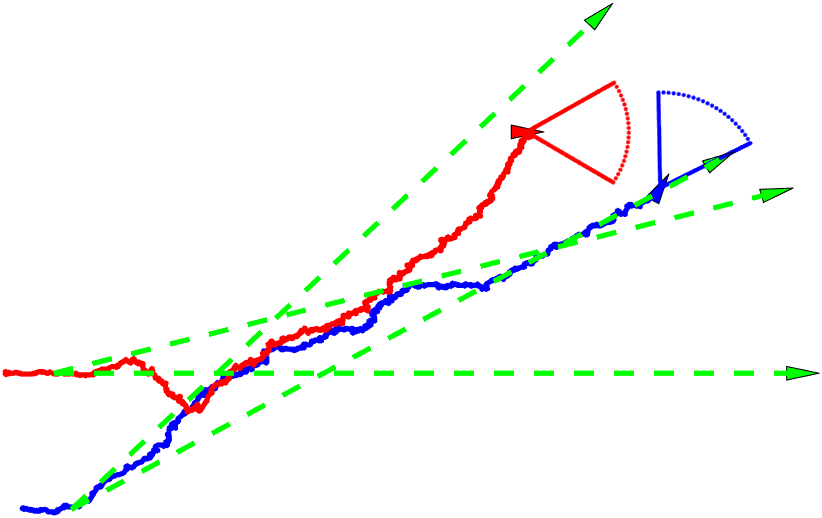} \\ 
             \vspace{3.2mm}
            \caption*{(e) {\smaller \sf BSG}: 2 robots and 4 targets.
            }
            \end{minipage}
            &
            \begin{minipage}{\advFigColWidth}%
            \vspace{2mm}
            \centering%
            \vspace{1mm}
            \includegraphics[width=\columnwidth]{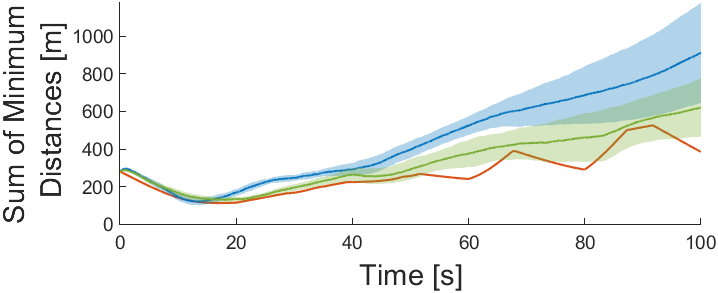} \\
            \vspace{-1mm}
            \caption*{(f) {\smaller \sf MetaBSG} vs. {\smaller \sf BSG} vs. {\smaller \sf Command}:
            2 robots and 4 targets.
            }
            \end{minipage}
            
            \end{tabular}
	\end{minipage} 
	\vspace{-3mm}
	\caption{\textbf{Multi-Robot Target Tracking with Untrustworthy External Commands: $2$ Robots Pursuing $2$ or $4$ Targets.} The robots select actions per {\smaller \sf MetaBSG}, per {\smaller \sf BSG}, or per the  {\smaller \sf External Command}, re-selecting actions with frequency $20$Hz. Particularly, the external command prescribes the robots with actions that follow desired trajectories, colored with magenta in (a) and (d). Across the cases, the targets traverse the same trajectories, which are predefined, \ie they are non-adaptive to the robots' motion. The command in (a) is suboptimal since it instructs the robots to be far from the targets; whereas, the command in (d) is near-optimal since it instructs the robots to intercept the targets.
 (a) and (d): The robots select actions per {\smaller \sf MetaBSG}. (b) and (e): The robots select actions per {\smaller \sf BSG}. (c)(f): Comparison of {\smaller \sf MetaBSG}'s, {\smaller \sf BSG}'s and {{\smaller \sf Command}}'s  \mbox{average effectiveness over $50$ Monte-Carlo trials.} 
	}\label{fig:command}
	\vspace{-9mm}
	\end{center}
\end{figure*}

\paragraph{Objective Function} The robots coordinate their actions $\{a_{i,\,t}\}_{i\myin \calN}$ to maximize an objective function at each time step $t$. This objective function is non-decreasing and submodular. The proof is presented in~\cite[Appendix III]{xu2023bandit}. 

\vspace{-4mm}
\begin{equation}\label{eq:distance}
   f_t(\,\{a_{i,\,t}\}_{i\myin \calN}\,)\,=\, \sum_{j\myin\calT}\left[-\sum_{i\myin \calN_j} \; \frac{1}{d_t(a_{i,\,t},\,j)}\right]^{-1},
\end{equation}where $\calN_j$ is the set of robots that can observe the target $j$, $d_t(a_{i,\,t},\,j)$ is the distance between robot $i$ and the estimated location  of target $j$.  Therefore, {$d_t(a_{i,\,t},\,j)$} is known only once robot $i$ has executed its action $a_{i,\,t}$ and the location estimate of target $j$ at time step $t$ has been computed.

By maximizing $f_t$, the robots aim to collaboratively to keep within their field of view the targets.  For example, when a robot $i$ achieves $0$ estimated distance from a target $j$, \ie when $d_t(a_{i,t},j)=0$, then indeed $\left[-\sum_{i\myin \calN_j} {1}/{d_t(a_{i,t},j)}\right]^{-1}=0$. On the other hand, when \underline{no} robot has target $j$ within its field of view, \ie when $\calN_j=\emptyset$, which is equivalent to target $j$ being infinitely far away from all robots, it is $\left[-\sum_{i\myin \calN_j} \; {1}/{d_t(a_{i,\,t},\,j)}\right]^{-1}=-\infty$ ---to account for the feasibility of our implementation, when $\calN_j=\emptyset$, we set $\left[-\sum_{i\myin \calN_j} \; {1}/{d_t(a_{i,\,t},\,j)}\right]^{-1}=-4d_{max}$, where $d_{max}$ is the largest sensing range among the robots. 

\paragraph{Performance Metric}
To measure how closely the robots track the targets, we consider a \textit{total minimum distance} metric.  We define the metric as the sum of the distances between each target and its nearest robot. 

\myParagraph{Computer System Specifications} We ran all simulations in MATLAB 2022b on a Windows laptop equipped with the Intel Core i7-10750H CPU @ 2.60 GHz and 16 GB RAM. 

\textbf{Code.} {{Our code will become available via a link herein.} \href{}{}}

\subsection{Evaluation of \sgmeta}\label{subsec:evaluation}

We evaluate \sgmeta in simulated target tracking scenarios where our external commands are either better or worse than \banalg with respect to our performance metric.

\myParagraph{Simulation Setup} 
We consider two scenarios of  target tracking: 
(i) $2$ robots with suboptimal commands vs.~$2$ targets, where the targets traverse straight lines with crossing (Fig.~\ref{fig:command}(a)--(c)); and (ii) $2$ robots with near-optimal commands vs.~$4$ targets, where two pairs of targets diverse with certain angles and traverse straight lines with crossing (Fig.~\ref{fig:command}(d)--(f)).  Each robot and target have different speeds, but we assume that all targets move with less speed than the robots.
In all scenarios, the robots re-select actions with frequency ($20$Hz). \mbox{We evaluate the algorithms across $50$ Monte-Carlo trials.}

\vspace{-0mm}
\myParagraph{Results} The simulation results are presented in Fig.~\ref{fig:command}. The following observations are due:
(i) The performance of \sgmeta approaches that of the optimal base learner in all scenarios (Fig.~\ref{fig:command}(c)(f)).  
In the case of $2$ robots vs.~$2$ targets (Fig.~\ref{fig:command}(a)(b)(c)), \sgmeta is actively close to \banalg when commands are inferior to \banalg (Fig.~\ref{fig:command}(a)(c)). In the case of $2$ robots vs.~$4$ targets (Fig.~\ref{fig:command}(d)(e)(f)), \sgmeta actively learns from superior commands and outperforms \banalg. In all scenarios, \sgmeta maintains near-optimal performance even if one of the base learners is not reliable.

\vspace{-1mm}

\section{Conclusion} \label{sec:con}
We introduced \sgmeta, the first algorithm for bandit submodular coordination in unpredictable and partially-observable environments with untrustworthy external commands. The commands are untrustworthy in that they offer \underline{no} performance guarantees; \eg they may be generated by human operators or machine learning algorithms. \sgmeta leverages a meta-algorithm to learn whether to follow the untrustworthy commands or a recently developed coordination algorithm that guarantees performance in unpredictable and partially-observable environments, namely, \banalg~\cite{xu2023bandit}.   \sgmeta asymptotically achieves the better performance out of the commands and \banalg, guaranteeing a bounded suboptimality against the optimal time-varying multi-robot actions. Thus, \textit{\sgmeta can also be interpreted as robustifying the untrustworthy commands.} We validated \sgmeta in simulated scenarios of target tracking.

\bibliographystyle{IEEEtran}
\bibliography{references}


\appendices

\section{Proof of \Cref{th:main}}
Substituting the approximation performance of \expthreeix, the external commands, and \banalg, we have 
{\small\begin{align}
&\mathbb{E}\left[\sum_{t=1}^{T}f_t(\,\calA_t^\sgmeta\,)\right] \nonumber\\
&\geq \max_{s\myin\{\human,\,\banalg\}}{\mathbb{E}\left[\sum_{t=1}^{T} f_t(\,\calA_t^s)\right]}  \nonumber\\
&\hspace{3.2cm} -O\left[\sqrt{T}\right] - \log{\left(\frac{1}{\delta}\right)}\,O\left[\sqrt{T}\right] \label{aux2:3} \\
&\geq \max\Bigg\{\frac{1}{2}\,\mathbb{E}\left[\sum_{t=1}^{T}f_t(\,\calA_t^\opt\,)\right] \nonumber\\
& -\tilde{O}\left[\sqrt{T\Delta(T)}\,\right] - \log{\left(\frac{1}{\delta}\right)}\,\tilde{O}\left[\sqrt{T}\right], \beta\,\mathbb{E}\left[\sum_{t=1}^{T}f_t(\,\calA_t^\opt\,)\right]\Bigg\}\nonumber\\
&  - O\left[\sqrt{T}\right] - \log{\left(\frac{1}{\delta}\right)}\,O\left[\sqrt{T}\right],\label{aux2:4}
\end{align}}with probability at least $1-\delta$, where \cref{aux2:3} holds from \cref{eq:bound_expthreeix}, and \cref{aux2:4} holds by substituting the bounds from \cref{eq:external_bound,aux:tracking_sec}. Hence, \cref{eq:main_bound} holds. \qed

\end{document}